# MCPI: Integrating Multimodal Data for Enhanced Prediction of Compound-Protein Interactions

Li Zhang, Wenhao Li, Haotian Guan, Zhiquan He, Mingjun Cheng, and Han Wang*

*Abstract*—The identification of compound-protein interactions (CPI) plays a critical role in drug screening, drug repurposing, and combination therapy studies. The effectiveness of CPI prediction relies heavily on the features extracted from both compounds and target proteins. While various prediction methods employ different feature combinations, both molecular-based and network-based models encounter the common obstacle of incomplete feature representations. Thus, a promising solution to this issue is to fully integrate all relevant CPI features. This study proposed a novel model named MCPI, which is designed to improve the prediction performance of CPI by integrating multiple sources of information, including the PPI network, CCI network, and structural features of CPI. The results of the study indicate that the MCPI model outperformed other existing methods for predicting CPI on public datasets. Furthermore, the study has practical implications for drug development, as the model was applied to search for potential inhibitors among FDA-approved drugs in response to the SARS-CoV-2 pandemic. The prediction results were then validated through the literature, suggesting that the MCPI model could be a useful tool for identifying potential drug candidates. Overall, this study has the potential to advance our understanding of CPI and guide drug development efforts.

## I. INTRODUCTION

IDENTIFYING Compound-Protein Interactions (CPIs) on a large scale is a critical step in drug discovery and development. By comprehensively understanding how a chemical compound interacts with various proteins in a living body, researchers can identify potential targets for drug development and better understand the mechanisms of action for existing drugs[1]. The identification of CPIs can also lead to the development of combination therapies[2, 3], where two or more drugs are used together to target multiple proteins or pathways involved in a disease. This approach can be especially useful in treating complex diseases such as cancer[4] or Alzheimer's disease  [5]. In addition, the identification of CPIs can also facilitate drug repositioning[6, 7], which involves repurposing existing drugs for new indications. By understanding the full range of proteins targeted by a drug, researchers can identify new therapeutic uses for the drug beyond its originally intended purpose. Finally, the identification of CPIs can also be useful in the modernization of traditional medicine, such as traditional Chinese medicine[8].

By identifying the specific proteins targeted by traditional remedies, researchers can better understand their mechanisms of action and potentially develop more effective and targeted treatments based on these traditional remedies. Overall, the identification of CPIs is a crucial step in the drug discovery process and has the potential to improve the efficacy and safety of drug treatments for a wide range of diseases.

CPI prediction using traditional biology-based methods and machine learning is challenging, while molecular docking[9] and molecular dynamics simulations[10] have been used for drug research for decades, identifying CPI from large-scale chemical spaces using current experimental methods is still difficult, especially for proteins with unknown structures[11]. Therefore, computational methods[12] have been increasingly applied to predict CPI.

Several computational methods have been proposed for predicting CPI, including tensor-product-based elements[13] between chemical substructures and protein families, supervised learning methods such as bipartite graphs[14], and feature selection techniques using support vector machines[15]. Traditional machine learning methods[16] have performed better in CPI prediction, but they require large-scale manual labeling and feature computation before modeling, which can be limiting when dealing with vast amounts of CPI data.

With the advancement of deep learning, various end-to-end frameworks have been introduced to overcome the inefficiencies of traditional machine learning methods[17-28]. Deep learning methods for predicting protein-compound interactions can be broadly categorized into two groups: those based on molecular structure data and those based on network data. The former utilizes protein and compound data represented as amino acid sequences and chemical structure formulas. Some of these methods, such as DeepDTA[29], WideDTA[30], and Conv-DTI[31], use convolutional neural

This work was supported by the Jilin Scientific and Technological Development Program (No. 20230201090GX, 20230401092YY, 20210101175JC), the Capital Construction Funds within the Jilin Province budget (grant 2022C043-2), and the Shenzhen Stability Support General Project (Type A) 20200826104014001. *(Corresponding author: Han Wang)*
Li Zhang is with School of Computer Science and Engineering, Changchun University of Technology, Changchun, China (e-mail: lizhang@ccut.edu.cn).
Wenhao Li is with School of Information Science and Technology, Institute of Computational Biology, Northeast Normal University, Changchun, China (e-mail: liwenhao0504@163.com).
Haotian Guan is with the College of Computer Science and Technology, Jilin University, Changchun, China (e-mail: guanht20@mails.jlu.edu.cn).
Zhiquan He is with Guangdong Key Laboratory of Intelligent Information Processing,Shenzhen,China (email: zhiquan@szu.edu.cn).
Mingjun Cheng is with School of Information Science and Technology, Institute of Computational Biology, Northeast Normal University, Changchun, China (email: chengmingjun@foxmail.com).
Han Wang is with School of Information Science and Technology, Institute of Computational Biology, Northeast Normal University, Changchun, China (email: wangh101@nenu.edu.cn).

networks to extract low-dimensional features of chemical compounds and proteins. Others, such as Gao et al.[32] and GraphDTA[33], use molecular graphs to represent chemical compounds and proteins by RDkit tool[34]. CPI-GNN[35], on the other hand, uses the r-radius subgraph algorithm[36] to obtain graph representations. In addition, Transformer CPI[37] , a novel Transformer neural network, was proposed, in which chemical compounds and proteins are considered as two sequences. Additionally, network-based methods have also been used for CPI prediction. Interaction networks are often used to represent interactions between molecules. Based on this idea, Zitnik et al.[38] proposed a deep learning method based on multimodal graphs to predict multi-drug side effects. Deep learning methods based on heterogeneous graphs, such as DTINet[39], deepDTnet[40] and NeoDTI[41], have been proposed to predict interactions between molecules. Moreover, some methods combine network and molecular information to predict drug targets. Yu et al.[42] proposed a method that combines network and molecular information for predicting drug targets.

This study proposed a new method called MCPI for predicting molecular Compound-Protein Interactions (CPI) by integrating relevant network and structure features. The network features were obtained from Protein-Protein Interaction (PPI) and Compound-Compound Interaction (CCI) separately to enhance the systemic background of those interactions. The molecular structures were represented using the distance matrix and Morgan fingerprint for the compound molecule and a pre-trained Word2vec model for the protein sequence representation. The features were learned using Gated CNN and ResNet deep learning models separately for molecular structures. Finally, a linear classifier was used to identify CPI using the embedded network and structure features. The experimental results showed that MCPI outperforms other methods on human and C. elegans benchmark datasets. The study also applied MCPI to COVID-19 data to find potential drugs for useful therapeutic leads in response to the SARS-CoV-2 pandemic.

II. MATERIALS AND METHODS

*A. Datasets*

In this study, we integrated both molecular features data and network features data to enhance the prediction performance of compound-protein interactions (CPI). The data for constructing the PPI and CCI networks contained both protein and compound interaction data. The protein interaction information was obtained from the STRING database[47], covering 24.58 million proteins from 14,094 species. When downloading protein interaction data from STRING, we filtered the data according to the reliability score of experimental data and set the threshold to 150.

Compound interaction data were obtained from the STITCH database[48]. For each compound, STITCH reliability-based score was created for all the interaction evidence. When downloading compound interaction data from STITCH, we filtered the data based on the reliability score by setting the threshold to 150 to remove the noisy data.

This experiment used two Benchmark datasets, Human and *C.elegans*, to assess model performance[49]. The positive samples in the datasets were retrieved from DrugBank[50] and Matador[51]. The negative sample candidates were collected by using a systematic screening framework and then filtered by feature scattering to obtain negative samples with high confidence[49]. The total number of samples in the Human dataset was 5502, and the *C.elegans* dataset was 6673. It contains the protein sequence and the compound SMILES. Single-atom molecules were removed during data preprocessing because they could not generate a distance matrix, and they represent a tiny proportion of the total sample. They do not affect the overall data distribution. In our experiments, the model was trained on the training set and then used the model on the test set to predict the results. As in Tsubaki et al.[35], the training, validation, and test sets were randomly split, and the ratio was 8:1:1.

*B. Model Architecture*

The MCPI model is a novel and integrated approach for predicting compound-protein interactions (CPI). By leveraging multiple sources of information, including the PPI network, CCI network, and structural features of CPI, the MCPI model effectively addresses the challenge of incomplete feature representations in existing prediction methods. It offers enhanced prediction performance compared to other methods. The model consists of four main components: Interaction network embedding, protein sequences and chemical compound coding, feature learning, and the linear classifier.

MCPI model uses protein sequences and compound SMILES as inputs to identify Compound-Protein Interactions. In the component of network embedding, the PPI and CCI networks are processed using Node2vec[52] to generate network feature vectors for each protein and compound. In the molecular structure coding part, the molecular structures are coded from the protein sequence and chemical compound. For compounds, the distance matrix and Morgan fingerprint are used to represent compound molecules by using the RDKit tool[34]. Then, the distance matrices were sent to the residual network for feature learning. After that, the feature vectors obtained from the residual network, the molecular fingerprints, and the network feature vectors of the CCI network were connected into a complete vector as a compound representation. For proteins, UniRef50 is used as a corpus and a pre-trained Word2Vec model[53] is employed to obtain protein sequence representation. The obtained feature vectors are then fed into a gated convolutional neural network[54] to learn high-level protein features. Finally, the feature vectors of proteins and compounds are separately fed into a fully connected layer, and a linear classifier is used to identify CPIs using the embedded network and structure features. Overall, the model architecture involves several state-of-the-art techniques for network representation learning, molecular structure coding, and deep learning. Fig. 1 presumably illustrates the model architecture.

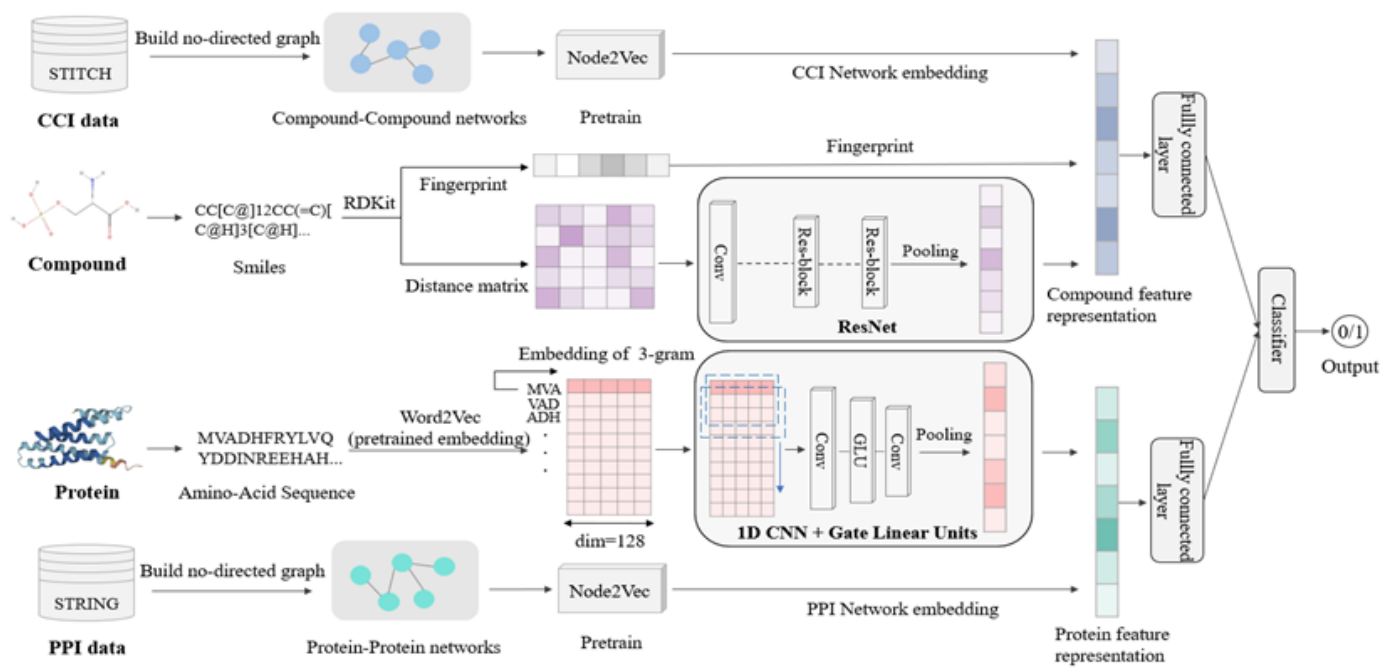

**Fig. 1.** MCPI model architecture. The PPI and CCI networks were processed using the Node2vec to generate network embeddings. The MCPI model combined the atom distance matrix with the molecular fingerprint to represent the compound molecule, and the pre-trained Word2vec model was applied for protein sequence representation. These features were fed into Gated CNN and ResNet to learn high-level molecular representations that capture the complex interactions between the compound and protein. Finally, a linear classifier is employed to identify the CPI.

*C. Features Extraction*

*1) Protein Sequence Coding:* The word embedding techniques, such as Word2vec, which were originally developed for natural language processing[55], have also been used to represent biological sequences, including DNA, RNA, and proteins. In MCPI, the Skip-Gram model in Word2vec[53] was used to obtain the protein sequence representation. Word2vec is an unsupervised machine learning technique that can learn high-quality representations of words by considering the context in which they appear. In the case of protein sequences, the Skip-Gram model in Word2vec has been commonly used to learn a distributed representation of the amino acid sequences. In Skip-Gram, the model learns to predict the context (neighboring amino acids) given a target amino acid., this allows the model to capture the syntactic and semantic relationships between amino acids. In our experiment, the large protein database UniRef50 was used as a corpus for pre-training. The amino acid sequences were considered as a sentence with different lengths. Then the original sequences were partitioned into overlapping 3-grams and trained using their contexts to obtain an embedding vector for each 3-gram. The dimensionality was set to 100. The feature vectors of the protein were delivered to the convolutional neural network to learn the high-level representation.

*2) Chemical Compound Coding:* Among previous studies, the molecular representations used in most approaches were divided into three main categories: Linear symbolic category, Molecular descriptor category, and Graphical symbol categories. Different representations had a significant impact on prediction performance. For example, a recent benchmark study showed that traditional descriptor-based machine learning models outperformed graph-based neural networks on 11 datasets relevant to drug discovery[46]. It's important to note that the best choice of molecular representation depends on the specific task at hand and the available data.

To represent the structural information of compound molecules, MCPI introduced two representations: the distance matrix and the Morgan fingerprint. The distance matrix is a matrix that represents the pairwise distances between all atoms in a molecule. Specifically, for a molecule with N atoms, the distance matrix is an $N \times N$ matrix, where each element $(i,j)$ represents the Euclidean distance between atoms i and j. Like Qian et al.[56], we used the RDKit[34] software to obtain distance matrices. To learn high-level representations from the distance matrices, MCPI used a residual network, which is a deep neural network architecture that allows for the training of very deep networks without suffering from vanishing gradients. The feature vectors output from the residual network were then concatenated with the Morgan fingerprint to form the final compound representation. This representation was then fed into a neural network to predict CPI.

*3) Compound/Protein Network Embedding:* Network embedding is a technique used to represent nodes in a network as low-dimensional vectors while preserving the network structure and topology. In this case, the pre-trained Node2vec models were applied to the PPI and CCI networks generated from the STRING and STITCH databases. The Node2vec algorithm is a popular network embedding method that learns node representations by performing random walks on the network and optimizing a skip-gram model. This results in vectors that capture the local and global network properties of each node. The models were trained to generate low-dimensional vector representations of proteins and compounds that capture their interactions and molecular activities within the network.

The implementation of the Node2vec algorithm in this paper used the Node2vec (version 0.4.3) module from the Python lib, where the Node2Vec parameters were set using these values (Embedding dimension: 128; Number of nodes searched in one random wander: WALK_LENGTH=80; Number of random wanders for each node: NUM_WALK=10; the probability of revisiting a wandering node: P=1; the search speed and range: R=1; whether to reflect the graph weights: Weight_Key=Weight). Finally, one 128-dimensional feature vector was generated for each node in the network. The network feature vectors are connected to the corresponding protein/compound molecular feature vectors to obtain the final feature representation.

### D. Feature Learning Model

*1) Protein Sequence Feature Learning:* After the word2vec pre-training process, each amino acid in the sequence is represented by a word vector, and the entire sequence can be represented as a matrix of these vectors. This matrix can then be used as input to a convolutional neural network (CNN) to extract high-level features of the protein sequence. Previous mainstream approaches to language models have been based on RNN[57], Dauphin et al.[54] proposed a novel gating mechanism integrating CNN networks applied to language models witch is useful for tasks that involve processing sequential data, as they can perform parallel computation and speed up training. Additionally, the hierarchical structure of CNNs can simplify learning and mitigate the vanishing gradient problem that can arise in RNNs. In the context of protein sequence analysis, CNNs can help to identify patterns and features that are important for predicting protein structure or function.

Therefore, we used a gated convolutional network with Conv 1D and gated linear units [54] to perform feature learning on proteins. Unlike regular convolution, the gated convolution was divided into two parts. One part was the convolutional activation value, which differs from normal convolution in that it was not activated with Tanh but was directly linear together with a sigmoid operator. The other part was the gate value, which was obtained directly linearly. Next, the gating unit made the element-wise product of the value obtained directly linearly and the value obtained after sigmoid activation to get the result after convolution. Finally, the convolution and gating unit were combined into a residual block. The overall structure is shown in Fig. 2, and the hidden layer $h_0,\ldots,h_N$ was calculated according to (1).

$$h_n(X) = (X * W_1 + b) \oplus \sigma(X * W_2 + c) \quad (1)$$

Where $X \in \mathbb{R}^{l \times m_1}$ is the input of the $h_l$ layer, $W_1 \in \mathbb{R}^{l \times m_1 \times m_2}$, $W_2 \in \mathbb{R}^{m_2}$, and $t \in \mathbb{R}^{m_2}$ are the learning parameters, $n$ is the number of hidden layers, $l$ is the amino acid sequence length, $m_1$ and $m_2$ are the dimensions of the input and hidden features, $k$ is the block size, $\sigma$ is the sigmoid function, and $\oplus$ is the product of elements between matrices. The gated convolutional network's output was the protein sequence's final representation. In our experiments, $n$ is 3, $m_1$ is 100, $m_2$ is 128, and $k$ is 3. The final output of the network is a 100-dimensional vector of protein representations.

*2) Compound Feature Learning:* The compound distance matrixes were obtained from the open-source cheminformatics software RDKit, which were represented as $D \in \mathbb{R}^{d \times d}$, where $d$ denotes the number of atoms in a single compound molecule. After obtaining the distance matrices, residual networks[58] were used to learn molecular structure features. Residual networks are a type of neural network architecture that uses shortcut connections between layers to address the problem of vanishing gradients and network degradation. By adding shortcut connections between every two/three layers, the input and output through redundant network layers remain unchanged, which speeds up computation. The compound feature learning architecture is shown in Fig. 3.

In MCPI, the ResNet-V2 module was used for compound feature learning. Since the information transfer between nodes is very important during the propagation of the residual neural network, ResNet-V2 made a comprehensive improvement on the position of skip connection and activation in the residual module. In ResNet-V2, a new concept of "pre-activation" was proposed, which means that the activation function (Relu and BN, batch normalization) was placed before the weight layer. For the traditional "post-activation" method, the output of the first residual unit was regularized by the BN layer and a shortcut layer was immediately added, but the combined signal was non-regularized. This non-regularized signal was directly used as the input of the next residual unit, which may affect the effect. The "pre-activation" method had all inputs regularized before the weighting layer, and this design could improve the performance of ResNet. The original residual module can be expressed in the generic form of Equations (2), (3).

$$y_l = h(x_l) + F(x_l, W_l) \quad (2)$$
$$X_{l+1} = f(y_1) \quad (3)$$

Where $h(x)$ is the mapping function, $F$ is the residual function, called the residual mapping, which consists of two or three convolution operations. $W_l$ is the weight of the residual block, $x_l$ and $X_{l+1}$ are the input and output of the lth residual unit, respectively. In ResNet V1, f is the activation function.

The idea of ResNet V2 is that this identity mapping occurs not only in individual residual units but throughout the entire network. To do this, two conditions need to be satisfied, which are $h(x_l) = x_l$ and $f(y_l) = y_l$. If $f$ is also an identity mapping, then $X_l+1 \equiv y_l$. Then taking (3) into (2) yields (4).

$$x_{l+1} = x_l + F(x_l, W_l) \quad (4)$$

In this way, the deep network unit $x_{l+1}$ can be expressed as the sum of the shallow network unit $x_l$ and the residual unit $F(x_l, W_l)$, which is a great property in backpropagation, and the gradient of the deep network will be directly passed to the shallow network. The gradient decay problem is well controlled.

For the compound distance matrix $D \in \mathbb{R}^{d \times d}$, the size of the distance matrix differs from compound to compound because the number of atoms in each compound is different. Although the residual network does not require a fixed input size and it can produce feature maps of any size, after the residual network processing, the fully connected layer requires a fixed size input, and cropping causes the loss of information. Therefore, the problem of fixed size originates from the fully connected layer, which is the final stage of the network. To solve this problem, we used a spatial pyramid pooling layer[59] after the residual network, and the primary purpose was to generate fixed-size outputs for arbitrarily sized inputs. Spatial pyramid pooling achieved multi-scale input by splitting the feature map of arbitrary size into 16, 4, and 1 blocks according to a (4 × 4), (2 × 2), and (1 × 1) grid. Then, it used a different size of max-pooling operation for each grid. The structure is shown in Fig. 4. The grid size determined each pooling layer's window size and step size. The window size is accomplished by dynamic calculations $d/4$, $d/2$ and $d/1$. The input of the pyramid pooling layer is $e \in \mathbb{R}^{d \times d \times m}$, where m is the number of channels. So, after three pooling operations, the obtained feature dimensions were $p_1 \in \mathbb{R}^{16 \times m}$, $p_2 \in \mathbb{R}^{4 \times m}$ and $p_3 \in \mathbb{R}^{1 \times m}$. Finally, they were stitched together to form a fixed-length feature vector $p_{final} \in \mathbb{R}^{21 \times m}$, used as the final output of compound feature extraction.

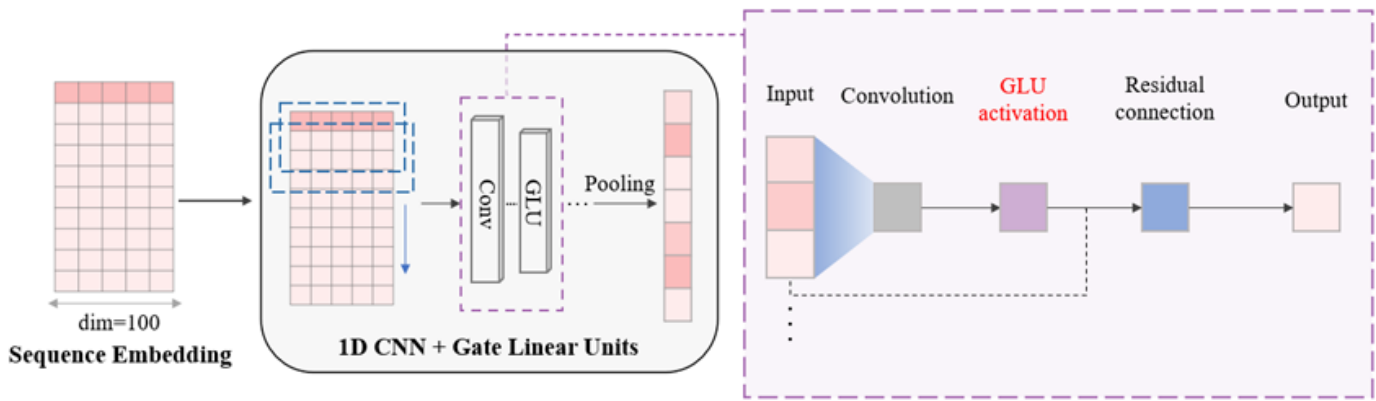

**Fig. 2.** Protein feature learning model architecture: The pre-trained word2vec model was used to obtain the representation of protein sequences. Each of the obtained word vectors was embedded into the continuous space. After getting the embedding matrices, they were fed into gated convolutional neural networks to learn high-level features.

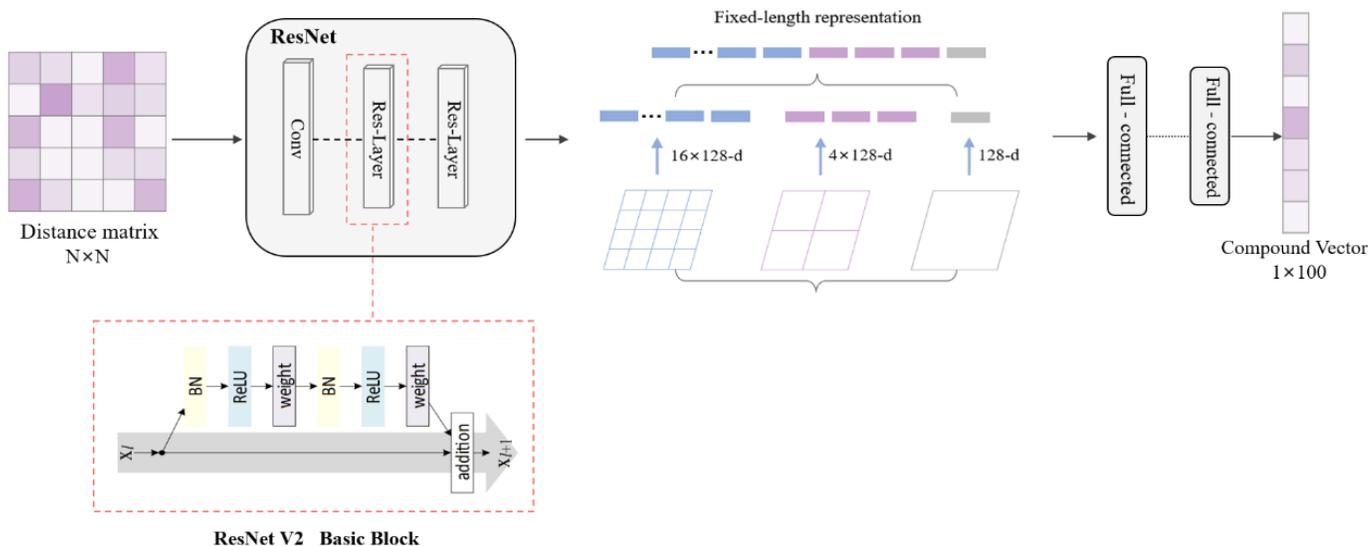

**Fig. 3.** Protein feature learning model architecture: The ResNet which combined by convolutional layers and res-layers was used to learn molecular structure features. Then the feature map of arbitrary size was sent to the spatial pyramid pooling layer to form a fixed-length feature vector.

*3) Linear Classifier:* After extracting the features of compounds and proteins, we obtain two 100-dimensional feature vectors: $p_{protein}$, which is the feature vector of the amino acid sequence processed by 1D-CNN, and $c_{compound}$, which is the feature vector of the compound distance matrix processed by residual network. Before inputting the feature vectors to the linear classifier, we need to connect the feature vectors from different data to obtain the complete features of compounds and proteins. We obtain the node embedding vectors of PPI network and CCI network using Node2vec, which are represented by $N_{protein}$ and $N_{compound}$, respectively. Additionally, we obtain the Morgan fingerprint vector of the compound, which is represented by $c_{finger}$. To create the protein expression, we add the two eigenvectors of protein, $p_{protein}$ and $N_{protein}$ to a vector $v_{protein}$. For the compound representation, we add the three characteristic vectors $c_{compound}$, $N_{compound}$, and $c_{finger}$ to the vector $v_{compound}$. Finally, we map the eigenvectors $v_{protein}$ and $v_{compound}$ to the same fixed dimensional potential space using the fully connected layer $f$. The calculation process was shown by (5), (6), (7).

$$v_{protein} = (p_{protein} \otimes N_{protein}) \quad (5)$$

$$v_{compound} = (c_{compound} \otimes c_{finger} \otimes N_{finger}) \quad (6)$$

$$T = f(v_{protein}) \otimes f(v_{compound}) \quad (7)$$

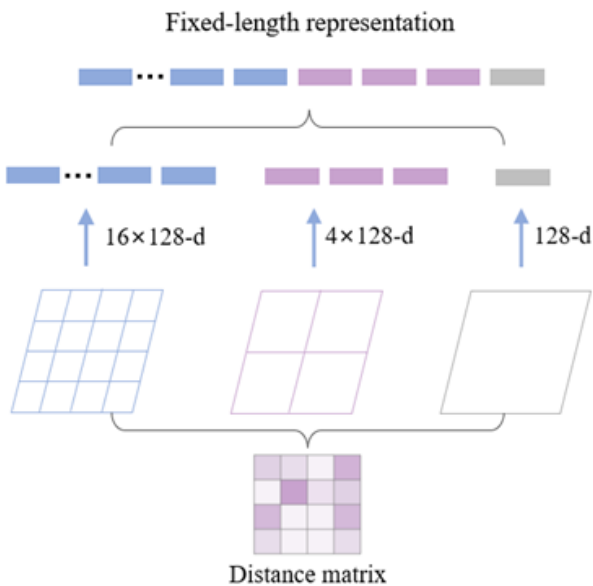

**Fig. 4.** Spatial pyramid pooling layer: First, for a feature map of arbitrary size, it is divided into blocks according to the grid of three different sizes. Then a different size of max-pooling operation is used for each grid. Finally, the vectors obtained from the pooling are stitched together to form a fixed-length feature vector.

The final vectors in our machine learning model were generated by performing an element product operation on protein and compound features of the same dimension. We then calculated the similarity between these vectors in the potential space, and fed the result into a fully connected layer to obtain the final prediction. If the prediction exceeded a predefined threshold (set at 0.5 by default), the model predicted an interaction between the input pairs. Otherwise, the model predicted no interaction. To optimize the model's parameters, we utilized the Adam algorithm[60], which updated the network weights more efficiently than the common stochastic gradient descent (SGD) method.

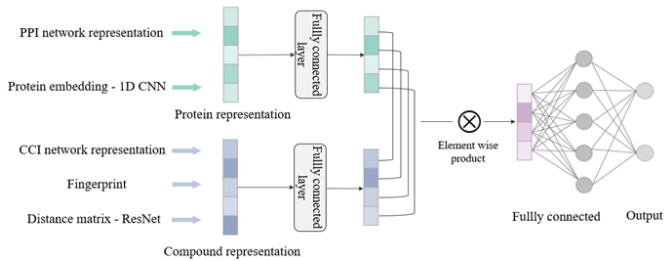

**Fig. 5.** Structure of the linear classifier: As shown in the figure, before feeding into the classifier, the embedding vectors of different features need to be connected to obtain the complete feature representations of compounds and proteins. After that, a fully connected layer was applied to map these feature vectors to the same latent space with a fixed dimension. Finally, the similarity of two vectors in the potential space was calculated by the element product operation, and then the fully connected layer was applied to obtain the final prediction result.

The MCPI updated the parameters by minimizing the loss function during the training process. We used binary cross-entropy as the loss function, which was defined as follows (8).

$$Loss = -\frac{1}{N}\sum_{i=1}^{N} y_i \cdot \log(p(y_i)) + (1 - y_i) \cdot \log(1 - p(y_i)) \quad (8)$$

Where the binary label, denoted as $y$, can take on either the value of 0 or 1. The probability that the model's output belongs to label $y$ is represented as $p(y)$, while $N$ signifies the total number of training samples. The binary cross-entropy is used to evaluate the performance of a binary classification model. Essentially, if the label $y$ is 1 and the prediction value $p(y)$ approaches 1, the loss function value should approach 0. Conversely, if the prediction value $p(y)$ approaches 0, the loss function value should be exceedingly high. Fig. 5 depicts the architecture of the classifier.

### III. RESULTS AND DISCUSSIONS

#### A. Experimental Environment and Parameters

We evaluated the predictive performance of MCPI by comparing it with state-of-the-art deep learning methods and traditional machine learning methods on two public benchmark sets, namely the human dataset and the *C.elegans* dataset[49]. To implement the model, we utilized Pytorch 1.8.0 for protein features, word2vec model from Gensim 4.0.1 to generate a 100-dimensional embedding vector of amino acid sequences, Node2Vec 0.4.0 to obtain 128 dimensional embedding vectors for network features, and Rdkit tool to obtain Morgan fingerprint vectors of length 50 for compound fingerprints. In the experiments, we set the learning rate to 0.0001 and the batch size to 1. We performed hyperparameter optimization by first setting the initial range of each hyperparameter, then randomly selecting parameter values for training, verifying the accuracy and narrowing the range of parameter values based on the prediction results until the best value was found.

#### B. Evaluation Metrics

During our experiments, we utilized four standard performance metrics for evaluating binary classification problems: the area under the receiver operating characteristic curve (AUC), the area under precision/recall curve(AUPR), Precision, and Recall. AUC is defined as the area under the ROC curve, and represents the likelihood of ranking the positive cases above the negative cases when the model is scored. Higher values of AUC, Precision, and Recall, which approach 1, indicate superior model performance. In addition, in order to highlight the resolution of the model for positive samples, we also introduced AUPR to evaluate the model. We calculated Precision and Recall using the (9), (10).

$$Recall = TP/(TN + FN) \quad (9)$$
$$Precision = TP/(TP + FP) \quad (10)$$

Where *TP* is the number of successfully identified compound-protein pairs with interactions (true positives). *FP* indicates the number of incorrectly identified interacting pairs that were predicted to be positive samples, but in fact, it was a negative sample (false positives). *FN* indicates the number of interacting pairs incorrectly identified as negative samples, that were predicted to be negative samples, but in fact, it was a positive sample (false negatives).

*C. Testing on Human Dataset*

We conducted a comparative analysis of MCPI with traditional machine learning methods and various advanced deep learning methods. The traditional machine learning methods we compared against included K-nearest neighbors[61], random forest[62], L2-logistic[63], and support vector machine (SVM)[64]. To ensure the robustness of our findings, we repeated all comparison experiments five times and calculated the mean and standard deviation of the results. We evaluated the experimental results using standard performance metrics such as the area under the receiver operating characteristic curve (AUC), the area under precision/recall curve(AUPR), accuracy (Precision), and recall (Recall). Table I presents the AUC, AUPR, Precision and Recall results for the human dataset in comparison with traditional machine learning methods.

Table I demonstrates that MCPI outperforms most traditional machine learning methods in all metrics, except for a slightly lower Precision performance than SVM. A comparison was also conducted with other deep learning-based methods, namely GraphDTA[33], GCN[65], GNN[35] and Transformer CPI[37]. GraphDTA, GCN, and GNN utilize molecular graphs to represent compound structural information and extract features using graph neural network models, while Transformer CPI modifies the Transformer architecture with a self-attentive mechanism to solve the sequence-based CPI classification task. Fig. 6 depicts a histogram with error lines, which indicates that MCPI significantly outperforms existing deep learning methods. Among these, Transformer CPI exhibit better performance with AUC values ranging from 0.971 to 0.975, while the molecular map-based method performs poorly compared to distance matrix-based and Transformer architecture-based methods. For human datasets, our MCPI method achieves the best results in terms of Precision, and Recall compared to the second best method, Transformer CPI, with performance gains of 4.80%, and 3.57%, respectively. The GCN-based method achieves an accuracy (Precision) of only 0.862, which is significantly lower than the deep learning and machine learning methods.

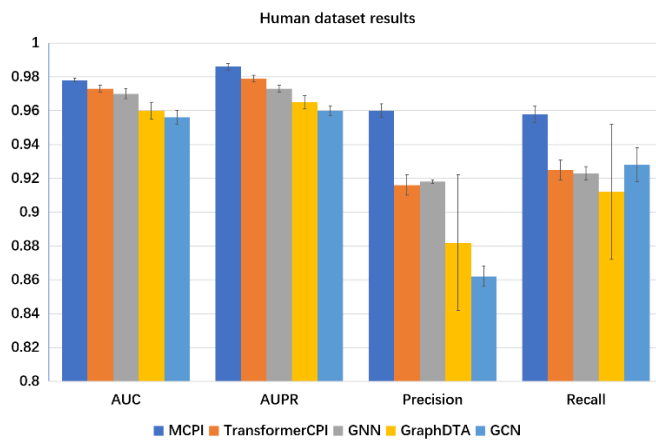

**Fig. 6.** Comparison with deep learning methods on the Human dataset, the figure shows the AUC, AUPR, precision, and recall scores of GraphDTA, GCN, GNN, TransformerCPI, and MCPI. The error line at the top of the histogram represents the standard deviation of the experiment, including the "upper error" and "lower error". As shown in the figure, MCPI outperforms other methods in all metrics

At the same time, in order to comprehensively evaluate the resolution of MCPI to positive samples, we used AUC and AUPR as evaluation indicators, the experimental results show that MCPI is superior to other methods in resolving positive samples.

*D. Testing on C.elegans Dataset*

Similarly, on the *C.elegans* dataset, we compare MCPI with traditional machine learning methods and several advanced deep learning methods. The AUC, AUPR, Precision, and Recall results for the human dataset compared to the traditional machine learning approach are shown in Table II.

As shown in Table II, MCPI performed better on C. elegans than on the Human dataset. Compared to traditional machine learning methods, MCPI significantly outperforms most traditional machine learning methods in all metrics, with an AUC of about 0.99. In addition, a comparison with other deep learning based methods was also conducted, which include GraphDTA, GCN, GNN and Transformer CPI. Fig. 7 shows the histogram with error lines.

TABLE I
HUMAN DATASET RESULTS

| Method | AUC | AUPR | Precision | Recall |
|---|---|---|---|---|
| Cover[43]KNN | 0.862 ±0.008 | 0.871 ±0.009 | 0.927 ±0.005 | 0.798 ±0.012 |
| Liaw[44]RF | 0.940 ±0.004 | 0.951 ±0.003 | 0.897 ±0.003 | 0.861 ±0.003 |
| Kleinbaum[45]L2 | 0.911 ±0.010 | 0.921 ±0.009 | 0.913 ±0.007 | 0.867 ±0.008 |
| Cortes[46]SVM | 0.910 ±0.023 | 0.923 ±0.026 | 0.910 ±0.023 | 0.939 ±0.025 |
| **MCPI** | **0.978 ±0.002** | **0.986 ±0.002** | **0.960 ±0.004** | **0.958 ±0.005** |

**Table I.** Human dataset results: This table shows the prediction performance of our model compared to traditional machine learning methods on the human dataset. MCPI significantly outperforms every traditional machine learning methods in all metrics.

TABLE II
*C.ELEGANS* DATASET RESULTS

| Method | AUC | AUPR | Precision | Recall |
|---|---|---|---|---|
| Cover[43]KNN | 0.858 ±0.008 | 0.862 ±0.009 | 0.801 ±0.010 | 0.827 ±0.009 |
| Liaw[44]RF | 0.902 ±0.007 | 0.896 ±0.005 | 0.821 ±0.006 | 0.844 ±0.007 |
| Kleinbaum[45]L2 | 0.892 ±0.004 | 0.903 ±0.005 | 0.890 ±0.003 | 0.877 ±0.004 |
| Cortes[46]SVM | 0.894 ±0.003 | 0.901 ±0.003 | 0.785 ±0.002 | 0.818 ±0.005 |
| **MCPI** | **0.990 ±0.002** | **0.991 ±0.004** | **0.955 ±0.005** | **0.954 ±0.004** |

**Table II.** *C.elegans* dataset results: This table shows the prediction performance of our model compared to traditional machine learning methods on the C. elegans dataset. MCPI significantly outperforms most traditional machine learning methods in all metrics.

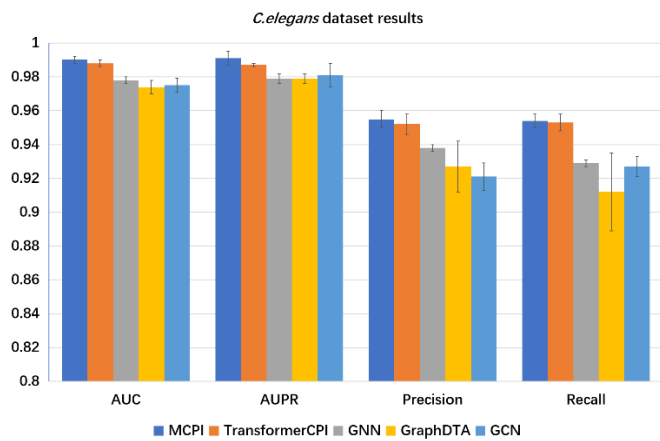

**Fig. 7.** Comparison with deep learning methods on the C. elegans dataset, the figure shows the AUC, precision, and recall scores of GraphDTA, GCN, GNN, TransformerCPI, and our proposed model. The error line at the top of the histogram represents the standard deviation of the experiment, including the "upper error" and "lower error. "As shown in the figure, MCPI outperforms other methods in all metrics.

As shown in Fig. 7, MCPI clearly outperforms existing deep learning methods. Among them, the method based on Transformer architecture has a better Recall of 0.953. When evaluated on *C.elegans* datasets, our MCPI method outperforms the second best method, Transformer CPI, across all key metrics including AUC, AUPR, Precision, and Recall, highlighting its superiority in accurately predicting CPIs. Overall, MCPI has more or fewer advantages over other methods in terms of AUC, Precision, Recall and is the best performance among all methods. Similarly, MCPI MCPI outperforms other advanced deep learning methods in the AUPR evaluation metric, demonstrating its superiority in both comprehensive CPI prediction performance and its ability to effectively distinguish positive samples.

*E. Ablation Experiments for Features*

In order to assess the efficacy of the MCPI model, which incorporates relevant network and structure features, we conducted a series of ablation experiments on both the Human dataset and C. elegans. The purpose of these experiments was to validate the effectiveness of each individual feature embedded in the model. To achieve this, we designed four different combination feature methods: the first method only included molecular features (Molecular features); the second method included PPI and CCI network features (Network features); the third method included both molecular and PPI network features (P features); and the fourth method included both molecular and CCI network features (C features). We compared the performance of these methods with that of the MCPI model and presented the experimental results in Table III and Table IV. The corresponding figures for the table data are shown in Fig. 8.

The results of the ablation experiments reveal a significant decrease in the model's performance. Notably, the MCPI model, which integrates the relevant network and structure features, provides the most informative data, with AUC values of 0.978 and 0.989 for the Human and *C.elegans* dataset.

TABLE III
HUMAN DATASET RESULTS

| Method | AUC | Precision | Recall |
|---|---|---|---|
| **MCPI** | **0.978** | **0.960** | **0.958** |
| P features[1] | 0.965 | 0.951 | 0.947 |
| C features[2] | 0.940 | 0.927 | 0.920 |
| Molecular features[3] | 0.934 | 0.921 | 0.913 |
| Network features[4] | 0.900 | 0.882 | 0.869 |

**Table III.** Comparison of four different feature combination methods with MCPI on Human dataset: the table shows the AUC, Precision, and Recall scores of P features method (containing only molecular and PPI networks features), C features method (containing only molecular and CCI networks features), Molecular features method (containing only molecular features), Network features method (containing only network features) and MCPI. The data in the table shows that the ablation process can significantly impair the performance of the model.

TABLE IV
*C.ELEGANS* DATASET RESULTS

| Method | AUC | Precision | Recall |
|---|---|---|---|
| **MCPI** | **0.990** | **0.955** | **0.954** |
| P features[1] | 0.981 | 0.946 | 0.942 |
| C features[2] | 0.952 | 0.924 | 0.917 |
| Molecular features[3] | 0.950 | 0.924 | 0.915 |
| Network features[4] | 0.883 | 0.870 | 0.859 |

**Table IV.** Comparison of four different feature combination methods with MCPI on C. elegans dataset: the table shows the AUC, Precision, and Recall scores of P features method (containing only molecular and PPI networks features), C features method (containing only molecular and CCI networks features), Molecular features method (containing only molecular features), Network features method (containing only network features) and MCPI. The data in the table shows that the ablation process can significantly impair the performance of the model.

However, intriguingly, the models that contained only single network features exhibited different performances. While neither the P features method nor the C features method performed as well as the MCPI model, the P features method appeared to outperform the C features method. Specifically, the PPI network features seem to provide more information than the CCI network features, especially in the C. elegans dataset where the performance gap between the two methods is larger. This may be due to the compounds on the C. elegans dataset being less connected and less correlated, providing less additional information for CPI prediction. The performance of the Molecular features method was intermediate between the MCPI and the individual Network features methods. In contrast, the Network features method exhibited a larger performance gap than the other models as it solely contained interaction network features without any molecular features.

Therefore, the molecular and interaction network features complement each other, and the effectiveness of the MCPI model with both features was also verified, demonstrating that the interaction network plays a crucial role in CPI prediction.

*F. Prediction of Potential Inhibitors for SARS-CoV-2 3CL Protease*

SARS-CoV-2, a member of the Coronaviridae family, is an enveloped virus with a positive single-stranded RNA genome.

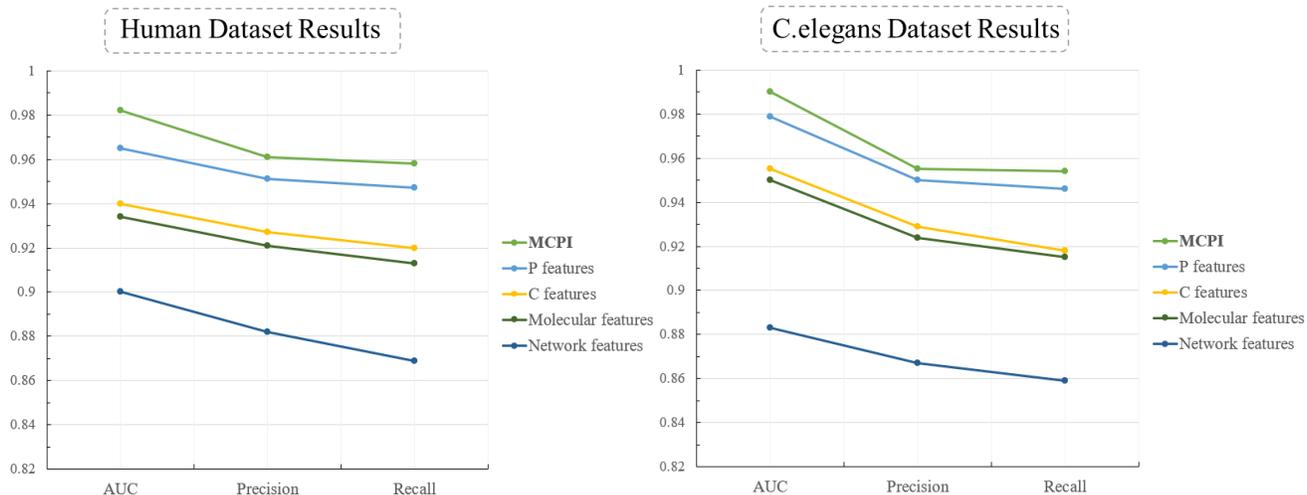
**Fig. 8.** Comparison of four different feature combination methods with MCPI on benchmark datasets.

This pathogen has led to almost 100 million cases of COVID-19 worldwide, resulting in millions of fatalities[66].

Although there have been some discoveries of treatments for COVID-19, their clinical efficacy is low or they need to be administered within a narrow therapeutic window. Therefore, there is a need for ongoing research into various treatment methods. SARS-CoV-2, like other coronaviruses, relies on an essential 3CL protease (3CLPro or MPRO) for processing its polyproteins. This enzyme plays a crucial role in the polyprotein processing of viral RNA translation and is considered the major protease. In fact, 3CLPro contains at least 11 cleavage sites on the large polyprotein 1AB (Replicase 1AB), and blocking its activity can prevent virus replication[67]. Hence, the potential of 3CL protease as a target for antiviral drugs has piqued the interest of scientists.

The SARS-CoV-2 pandemic necessitates the swift development of drugs, and repurposing existing drugs is a viable approach to overcome the research and development obstacles associated with creating new drugs. Repurposed drugs also carry lower risk of failure and require less investment than new drug development [68]. Therefore, this study aimed to utilize our predictive model to identify potential inhibitors among drugs already approved by the US Food and Drug Administration (FDA). Our model was trained on a dataset consisting of molecules screened for fragments bound to SARS-CoV-2 3CL protease (3CLpro) using crystallographic techniques. We chose the SARS-CoV-2 major protease with active site 6YB7 (PDB ID) as the predicted target based on data published by the Diamond Light Source group, which contains roughly 880 sample compounds with 78 interactions. Furthermore, we extracted all drugs in DrugBank [50] except those in the training dataset and generated test samples by combining each drug with the SARS-CoV-2 major protease (6YB7). Table V includes some of the compounds (drugs) predicted to interact with proteases.

Upon analyzing the prediction results from the literature, we discovered that anti-cancer drugs and anti-viral medications may have more interactions that could assist in treating the Corona Virus. For instance, Ibrutinib, an anti-neoplastic drug used to treat chronic lymphocytic leukemia, is a tyrosine kinase (BTK) inhibitor and has been investigated as a possible target for reducing the overly severe immune response in Corona Virus [69]. Similarly, Imatinib, another tyrosine kinase (BTK) inhibitor used to treat leukemia, has been proven to effectively treat SARS-CoV-2 infection, according to related studies [70]. Famotidine, a competitive histamine-2 (H2) receptor antagonist, is also a promising drug, slowing disease progression in Corona Virus patients by reducing the histamine-mediated cytokine storm [71]. Furthermore, Linagliptin, a DPP-4 inhibitor used to treat type II diabetes, can reduce Corona Virus severity by reducing inflammation [72]. Although the following drugs are predicted to be potentially effective, they are not currently known to have inhibitory effects on Corona Virus: Mechlorethamine, a nitrogen mustard compound and anti-tumor drug; Abacavir, an anti-viral nucleoside reverse transcriptase inhibitor used to treat HIV and may be effective against Corona Virus; Oxaprozin, an anti-inflammatory drug mainly used for treating arthritis; and Sulfadiazine, a sulfonamide compound anti-bacterial drug used to treat various bacterial infections, such as bronchitis, which we speculate may be effective against bronchitis caused by Corona Virus.

TABLE V
PREDICTDE POTENTIAL DRUGS FOR COVID-19

| ChEMBL ID | Drug Name | Evidence |
|---|---|---|
| ChEMBL427 | Mechlorethamine | - |
| ChEMBL1380 | Abacavir | - |
| ChEMBL902 | Famotidine | Hogan[50][52] |
| ChEMBL880 | Famciclovir | - |
| ChEMBL2105395 | Ibrutinib | Cherian[49][51] |
| ChEMBL446 | Sulfadiazine | - |
| ChEMBL1071 | Oxaprozin | - |
| ChEMBL941 | Imatinib | Morales[51][53] |
| ChEMBL1577 | Methyclothiazide | - |
| ChEMBL237500 | Linagliptin | Katsiki[52][54] |

In this experiment, we identified potential inhibitors of SARS-CoV-2 3CLPro, and these candidates showed similarities in both chemical and pharmacological classification. Moreover, MCPI also predicted compounds that have shown efficacy in some ongoing studies, making this a significant finding. Nevertheless, proper drug application necessitates in vitro and in vivo validation experiments, as well as clinical trials, to verify the drug's efficacy and other desired properties.

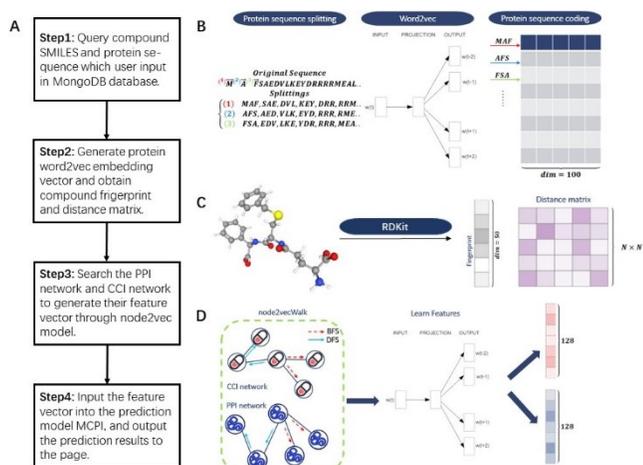

**Fig. 9.** MCPI server overview. (A) Main procedures; (B) Generate word2vec vector; (C) Use RDKit to obtain the fingerprint and distance of the compound_matrix; (D) The feature vector is generated through Node2vec.

## IV. ACCESSIBLE APPLICATION

We have developed an accessible MCPI server for researchers which enables us to predict potential interactions between drug molecules and target proteins, based on molecular features and network features of drug data and omics-scale protein data. Users simply need to provide the protein sequence and compound SMILES as inputs to MCPI. During the calculation process, MCPI can leverage additional data, including the compound distance matrix, compound molecular fingerprint, and biological interaction network, based on the user's input. This comprehensive approach ensures accurate predictions of compound-protein interactions.

The overall architecture of the MCPI server is depicted in Fig. 9. Upon user input, the server generates word2vec encoding for protein residue sequences and utilize key functions of RDKit to obtain the fingerprint and distance matrix for compounds. Subsequently, the MCPI server searches the protein and compound interaction networks, employing the encapsulated Node2vec model to obtain network encoding. Finally, the acquired feature code is inputted into the prediction model, and the results are outputted on the page. The web server is available at http://47.99.71.176:5000/index, respectively.

## V. CONCLUSION

In this work, we introduced MCPI, a novel prediction model that aims to improve the accuracy of predicting protein-compound interactions. To achieve this, we integrated the PPI network, CCI network, and structural features of molecular CPI. To extract protein features, we utilized a gated CNN that operates on amino acid sequences pre-trained by Word2vec. For compounds, we used a combination of distance matrices and fingerprints, which allows for a comprehensive representation of both the molecular structure and semantic features. We further enhanced our model by using residual networks to extract advanced molecular features that outperform other methods that rely solely on pre-trained embeddings. To capture network embeddings, we used the Node2vec model on the interaction network. Our approach outperforms previously proposed CPI and traditional machine learning-based models, as demonstrated through our evaluation. Additionally, we leveraged MCPI to identify potential inhibitors among FDA-approved drugs for SARS-CoV-2, validating our predictions against literature. Our work may provide valuable insights for drug development and may guide future efforts in this area.